\documentclass[aps,prl,twocolumn,floatfix,amsmath,amssymb,superscriptaddress]{revtex4-2}
\usepackage[final]{graphicx}
\usepackage{bm}
\usepackage{afterpage}
\usepackage{color}
\usepackage{comment}
\usepackage[colorlinks=true,urlcolor=blue,linkcolor=blue,citecolor=blue]{hyperref}

\newcommand{\beginsupplement}{%
        \setcounter{table}{0}
        \renewcommand{\thetable}{S\arabic{table}}%
        \setcounter{figure}{0}
        \renewcommand{\thefigure}{S\arabic{figure}}%
}

\begin{document}

\title{Emergent Antiferromagnetism in a $Y$-Shaped Kekul\'e Graphene}

\author{Chenyue Wen}
\affiliation{School of Physics, Beihang University,
Beijing, 100191, China}
\affiliation{Fert Beijing Institute, MIIT Key Laboratory of Spintronics, School of Integrated Circuit Science and Engineering,
Beihang University, Beijing 100191, China}

\author{Wanpeng Han}
\affiliation{School of Physics, Beihang University,
Beijing, 100191, China}

\author{Xukun Feng}
\affiliation{Research Laboratory for Quantum Materials, Singapore University of Technology and Design, Singapore 487372, Singapore}

\author{Xingchuan Zhu}
\affiliation{Interdisciplinary Center for Fundamental and Frontier Sciences, Nanjing University of Science and Technology, Jiangyin, Jiangsu 214443, P. R. China}

\author{Weisheng Zhao}
\affiliation{Fert Beijing Institute, MIIT Key Laboratory of Spintronics, School of Integrated Circuit Science and Engineering,
Beihang University, Beijing 100191, China}

\author{Shengyuan A. Yang}
\affiliation{Research Laboratory for Quantum Materials, Singapore University of Technology and Design, Singapore 487372, Singapore}

\author{Shiping Feng}
\affiliation{ Department of Physics,  Beijing Normal University, Beijing, 100875, China}

\author{Huaiming Guo}
\email{hmguo@buaa.edu.cn}
\affiliation{School of Physics, Beihang University,
Beijing, 100191, China}

\begin{abstract}
Antiferromagnetic (AF) transitions of birefringent Dirac fermions created by a $Y$-shaped Kekul\'e distortion in graphene are investigated by the mean-field theory and the determinant quantum Monte Carlo simulations. We show that the quantum critical point can be continuously tuned by the bond-modulation strength, and the universality of the quantum criticality remains in the Gross-Neveu-Heisenberg class. The critical interaction scales with the geometric average of the two velocities of the birefringent Dirac cones, and decreases monotonically between the uniform and the completely depleted limits. Since the AF critical interaction can be tuned to very small values, antiferromagnetism may emerge automatically, realizing the long-sought magnetism in graphene. These results enrich our understanding of the semimetal-AF transitions in Dirac-fermion systems, and open a new route to achieve magnetism in the graphene.
\end{abstract}


\maketitle

\paragraph{{\color{blue}Introduction.}}
Graphene features linear dispersions near the Fermi energy~\cite{geim2010rise,RevModPhys.81.109}, which is described by a massless Dirac equation. The electrons therein are thus called Dirac fermions, which have been the origin of various exotic properties~\cite{novoselov2005two,zhang2005experimental,katsnelson2006chiral}. While theoretical studies have suggested abundant topological or ordered phases in graphene, as generated by various kinds of mechanisms~\cite{PhysRevLett.101.087204,PhysRevB.80.113102,PhysRevLett.95.226801,PhysRevLett.98.186809,PhysRevLett.100.110405,PhysRevLett.107.066801,PhysRevLett.126.206804}, their experimental observations are lacking due to the overall weak correlation effects in graphene.

Achieving magnetism in graphene is actively pursued in the hope of its spintronic applications. Although pristine graphene has a non-magnetic ground state, magnetic orders have been observed at the zigzag edges, defects, and hydrogen-terminated vacancies~\cite{slota2018magnetic,PhysRevLett.100.047209,PhysRevLett.99.177204,PhysRevB.99.115442,PhysRevB.90.035403,PhysRevB.75.125408,gonzalez2016atomic,yazyev2010emergence}. The emergent magnetism is generally related to the associated localized states~\cite{PhysRevLett.102.096801}, for which the correlation effect is greatly enhanced, resulting in magnetic ordering at much weaker interactions~\cite{PhysRevB.87.155441,PhysRevLett.106.226401,PhysRevB.81.115416}. From this physical mechanism, it is clear that the above magnetisms are restricted to specific regions of graphene, e.g., edges or defects. Up to now, inducing a robust long-range magnetic order in the bulk of graphene remains a challenge.

In this work, we propose that the $Y$-shaped Kekul\'e distortion can induce a global antiferromagnetic (AF) order in graphene. Our work is motivated by the recent experiments which successfully realized such distortion in graphene grown on Cu(111)~\cite{gutierrez2016imaging} and on transition metal dichalcogenides substrates~\cite{helin2022}.
We show that the $Y$-shaped bond texture modifies the Fermi velocities of two pairs of low-energy bands differently, creating birefringent Dirac fermions. AF transitions of such fermions are investigated by two complementary methods: the mean-field theory and the large-scale determinant quantum Monte Carlo (DQMC) simulations.  Both methods predict a SM-AF transition, and find that the critical interaction is proportional to the geometric average of the two velocities of birefringent Dirac fermions. The DQMC simulations quantitatively determine the critical values using finite-size scaling. Besides, it is found that the quantum criticality remains in the Gross-Neveu-Heisenberg universality class. Our results show that the $Y$-shaped distortion enables an additional degree of control of Dirac fermions, and provides a new way to tune the quantum critical point. Importantly, we reveal that when the distortion strength is large enough, the critical interaction may be well below the actual value of $U/t$ in graphene~\cite{mynote1,tang2018role}, generating global AF long-range order. Considering the recent progress in engineering the $Y$-shaped Kekul\'e distortion~\cite{gutierrez2016imaging,helin2022}, our result provides a feasible approach to realize magnetic graphene in experiment.

\paragraph{\color{blue}Model.}

\begin{figure}[htbp]
\centering \includegraphics[width=7.5cm]{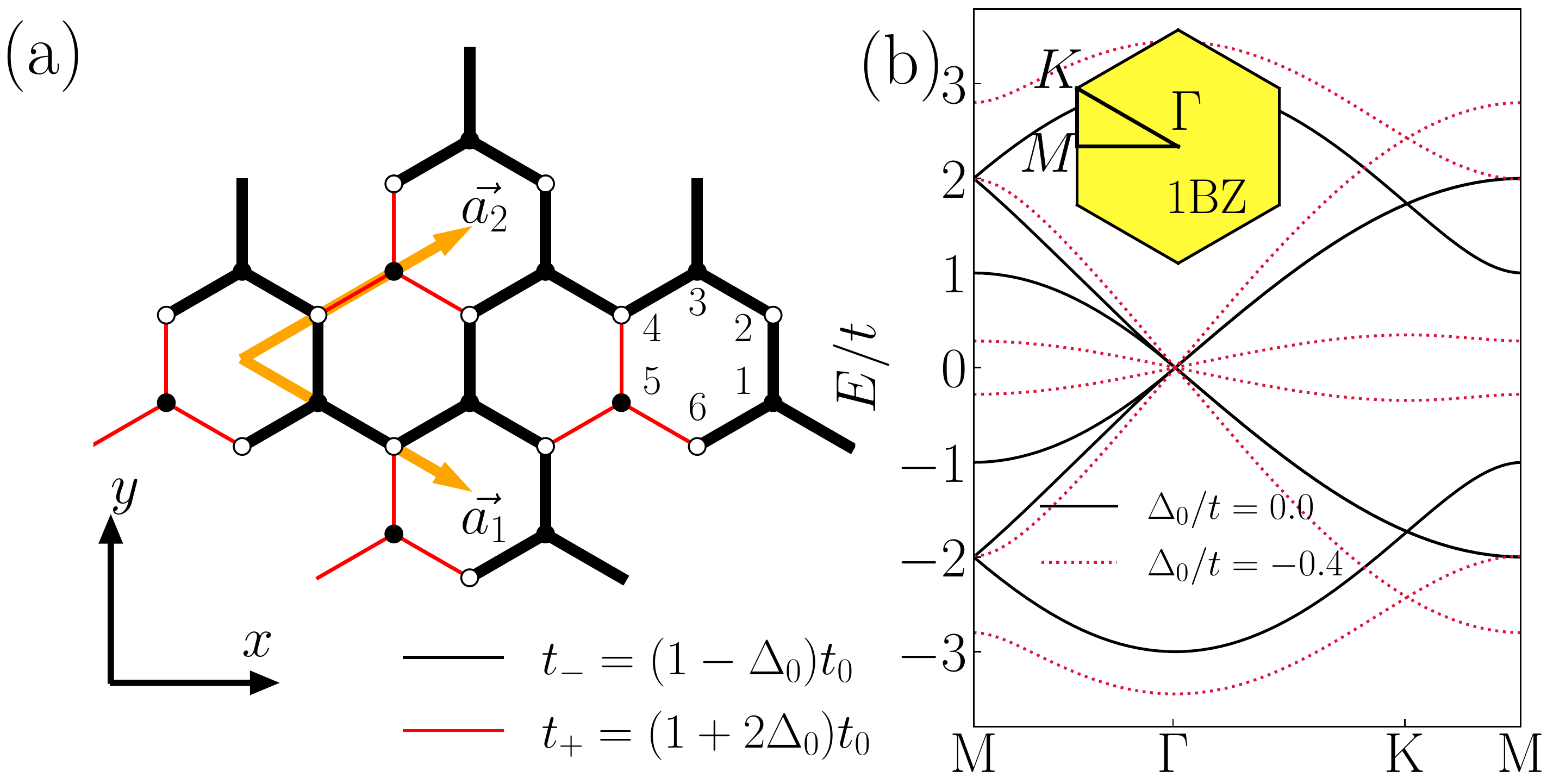} \caption{(a) Honeycomb lattice with a $Y$-shaped bond modulation. Black and white dots label the two sublattices, and black and red lines distinguish different
bond strengths. The different sites in the unit cell are marked by a set of integers ($1-6$). (b) The energy spectrum along the high-symmetry directions of the Brillouin zone at $\Delta_0/t=0$ and $-0.4$. Inset of (b) is the $1$st Brillouin zone with the high-symmetry points labeled.}
\label{fig1}
\end{figure}

Our study is based on the Hubbard model defined on a honeycomb lattice with a $Y$-shaped Kekul\'{e} distortion:
\begin{align}\label{eq1}
H=-\sum_{\langle ij\rangle \sigma}t_{ij}(c^{\dagger}_{i\sigma}c_{j\sigma}+\textrm{H.c.})+U\sum_{i}n_{i\uparrow}n_{i\downarrow},
\end{align}
where $c_{i \sigma}^{\dagger}$ and $c_{i \sigma}$ are the creation and annihilation operators, respectively, at site $i$ with spin $\sigma=\uparrow, \downarrow$; $\langle ij\rangle$ denotes nearest neighbors; $n_{i \sigma}=c_{i \sigma}^{\dagger} c_{i \sigma}$ is the number operator of electrons; and $U$ is the on-site repulsion. Due to the $Y$-shaped modulation, the hopping amplitudes are modified as $t_{ij}=(1+2\Delta)t$ or $(1-\Delta)t$, depending on the position and direction of the bond (see Fig.~1(a)), where $\Delta\in[-0.5, 1]$. The modifications are defined in such a way to keep the total band width more or less unchanged. The pristine honeycomb lattice is restored at $\Delta=0$. In the limit $\Delta=-0.5$, the $5$th site in each unit cell is completely isolated from the lattice (see Fig.~1(a)), generating a $1/6$-depleted honeycomb lattice. This is similar to the Lieb lattice (also known as $1/4$ depleted square lattice)~\cite{PhysRevLett.62.1201}. $\Delta>0$ interchanges the strong and weak bonds as in the $\Delta<0$ case, and the lattice is completely broken into isolated four-site stars at $\Delta=1$.

Under the $Y$-shaped distortion, the lattice has a six-site unit cell (see Fig.~\ref{fig1}(a)).
Let's first neglect the interaction term in (\ref{eq1}), then
the electronic band structure contains six dispersive bands, as shown in Fig.~1(b). By a Brillouin zone (BZ) folding process, the bands of the $\Delta=0$ case in such a plot can be obtained from the pristine graphene band structure, where the two inequivalent Dirac points at the corners of the BZ are folded to the $\Gamma$ point and the Dirac cones will coincide. For $\Delta \neq 0$, although the linear dispersions remain near the Dirac points, the Dirac cone degeneracy is lifted, generating birefringent Dirac fermions with two different Fermi velocities, as in Fig.~1(b). By projecting the full Hamiltonian to the low-energy space at $\Gamma$, an effective Hamiltonian can be deduced~\cite{mynote}. The energy spectrum contains four branches with $E=\pm \frac{3}{2}(1-\Delta)tk$ and $E= \pm \frac{3}{2}\alpha(1-\Delta)tk$, with $\alpha=\frac{1+2\Delta}{\sqrt{1+2\Delta^2}}$. The two Fermi velocities are respectively given by
\begin{equation}
  v_1=\frac{3}{2}(1-\Delta)t,\qquad v_2=\frac{3}{2}\alpha(1-\Delta)t.
\end{equation}
It is noted that $v_2$ becomes zero in two limits: $\Delta=-0.5$ or 1, generating two or four flat bands at the Fermi energy. 
When $\Delta$ approaches the above two limits, the low-energy bands will be flattened, and the corresponding states tend to localize on the sites connected by the weakened bonds. Then, the correlation effect in the low-energy bands will be enhanced, and we expect that the AF critical interaction will be significantly reduced. This expectation will be explored by two theoretical approaches, as we demonstrate below.


\paragraph{\color{blue}Mean-field theory approach.}

\begin{figure}[htbp]
\centering \includegraphics[width=8.5cm]{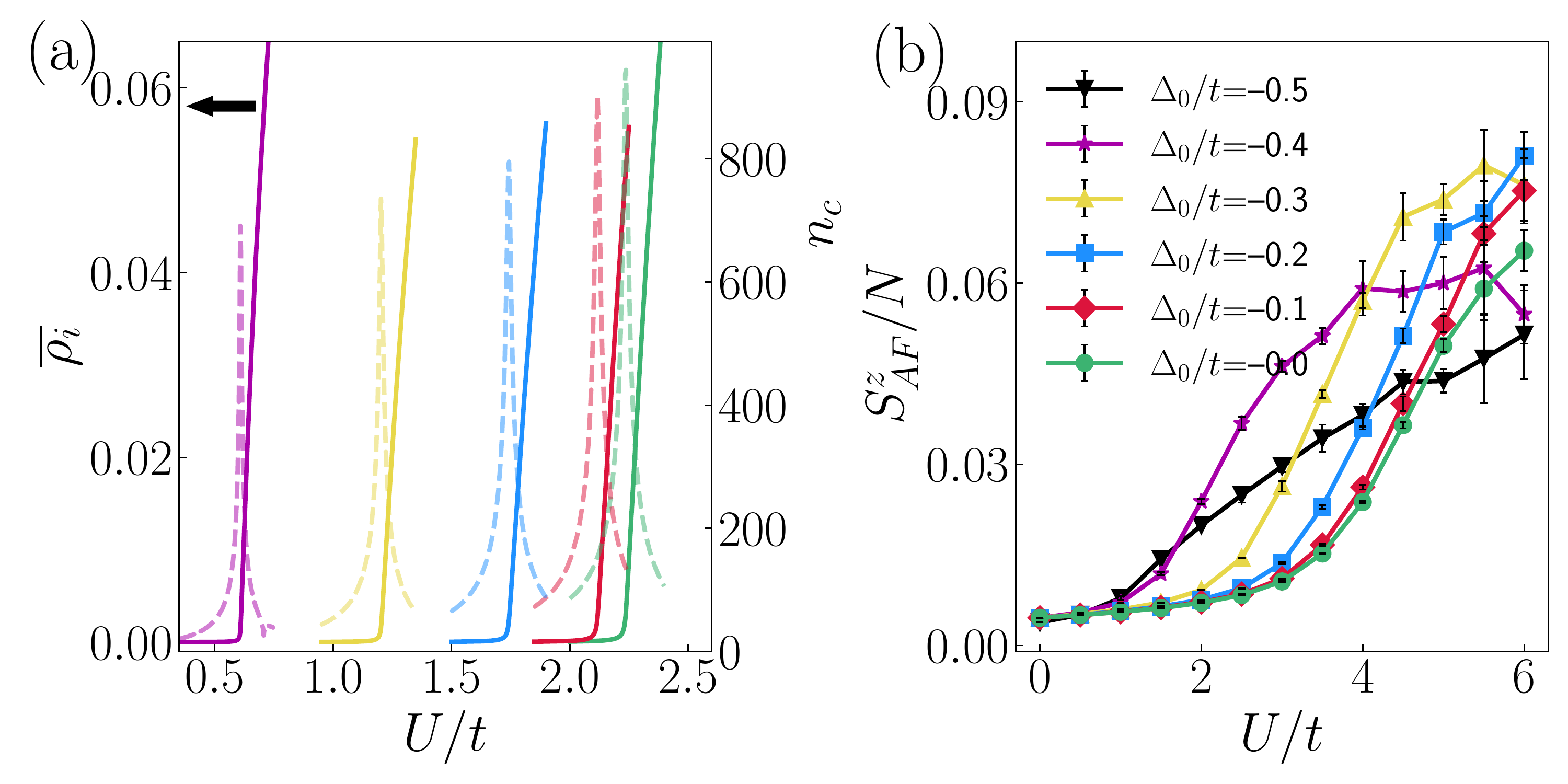} \caption{(a) The mean-field order parameter and the cycling time $n_c$ to get convergent solution in the self-consistent process as a function of $U$ for various values of $\Delta_0$. The sharp peak in the $n_c$ curve can steadily determine the transition point. (b) The AF structure factor obtained by DQMC as a function of $U$ on a lattice with the size $L=6$.}
\label{fig2}
\end{figure}

Within the mean-field approximation, the interacting term $Un_{i \uparrow} n_{i \downarrow}$ is decoupled in the density channel as~\cite{PhysRevB.102.155152,PhysRevB.97.235152,Zhu_2019} $n_{i \uparrow} n_{i \downarrow} \approx n_{i \uparrow}\left\langle n_{i \downarrow}\right\rangle+\left\langle n_{i \uparrow}\right\rangle n_{i \downarrow}-\left\langle n_{i \uparrow}\right\rangle\left\langle n_{i \downarrow}\right\rangle$.
To incorporate the possible AF order, the averages of the operators are written as
\begin{equation}
  \langle n_{i \uparrow(\downarrow)}\rangle=\frac{1}{2}\pm \rho_i,\qquad \text{for } i\in\{1,3,5\},
\end{equation}
and
\begin{equation}
  \langle n_{i \uparrow(\downarrow)}\rangle=\frac{1}{2}\mp \rho_i,\qquad \text{for } i\in\{2,4,6\},
\end{equation}
with $\rho_i$ being the order parameters, where $i$ labels the six sites in a unit cell. We consider equal number of spin-up and spin-down electrons, such that the following restriction applies on the six order parameters: $\rho_1+\rho_3+\rho_5=\rho_2+\rho_4+\rho_6$, which means only five of them are independent.

In the momentum space, the mean-field Hamiltonian is then
\begin{equation}
  H_\text{MF}=\sum_{\bm k\sigma}\psi_{{\bm k}\sigma}^{\dagger}{\cal H}^{\sigma}({\bm k})\psi_{{\bm k}\sigma}+E_0
\end{equation}
with
\begin{equation}
  {\cal H}^{\sigma}({\bm k})=\left(\begin{array}{cc}
h^{\sigma}_{11} & h_{12} ({\bm k}) \\
h^{\dagger}_{12}({\bm k}) & h^{\sigma}_{22}  \\
\end{array}\right),
\end{equation}
where
$\psi_{{\bm k}\sigma}=(c_{1,{\bm k}\sigma},\cdots,c_{6,{\bm k}\sigma})^{T}$ is a six-element basis, $h^{\sigma}_{11}=\mp\textrm{diag}(\rho_1,\rho_3,\rho_5)$, $h^{\sigma}_{22}=\pm\textrm{diag}(\rho_2,\rho_4,\rho_6)$, and the constant $E_0=NU/4+NU\sum_{i=1}^{6}\rho_i^2/6$ with $N$ the total number of sites. The order parameters $\rho_i$ in the ground state can be calculated by minimizing the total energy from the mean field Hamiltonian.

Although the order parameters $\rho_i$ are generally different within the unit cell due to the inhomogeneous bond textures, their curves as a function of $U$ have the same features. Hence, we use their arithmetic mean,  $\overline{\rho}=\frac{1}{6}\sum_{i=1}^{6}\rho_i$, to characterize the AF transition.
Figure~2(a) shows $\overline{\rho}$ as a function of $U$ for various values of $\Delta$ at zero temperature. The values of $\overline{\rho}$ keep almost zero at weak interactions, and the system is in the semimetal phase. After passing a critical interaction $U_c$, $\overline{\rho}$ becomes finite, and increases rapidly with $U$, suggesting the AF order develops in the system. As the absolute value of $\Delta$ increases, the curves move leftward, indicating that the critical interaction decreases monotonically with increasing $\Delta$.

Since the curves of $\overline{\rho}$ is continuous, it is not straightforward to determine $U_c$. Here, we use the critical slowing down at the phase transition to determine $U_c$. Specifically, the self-consistent process becomes the slowest at the transition point, which is reflected in the maximum cycling time $n_c$ to get convergent self-consistent results. We checked that the critical value at $\Delta=0$ obtained by this method is in good agreement with the previous study~\cite{Sorella_1992}.
Our result is shown in Fig.~3. One finds that the critical interaction of the semimetal-AF transition decreases monotonically from $U_c/t=2.23$ at $\Delta=0$ to $U_c/t=0$ at $\Delta=-0.5$. Similarly, $U_c$ also decreases monotonically with $\Delta$ for  $\Delta>0$~\cite{mynote}.

\begin{figure}[t]
\centering \includegraphics[width=8.5cm]{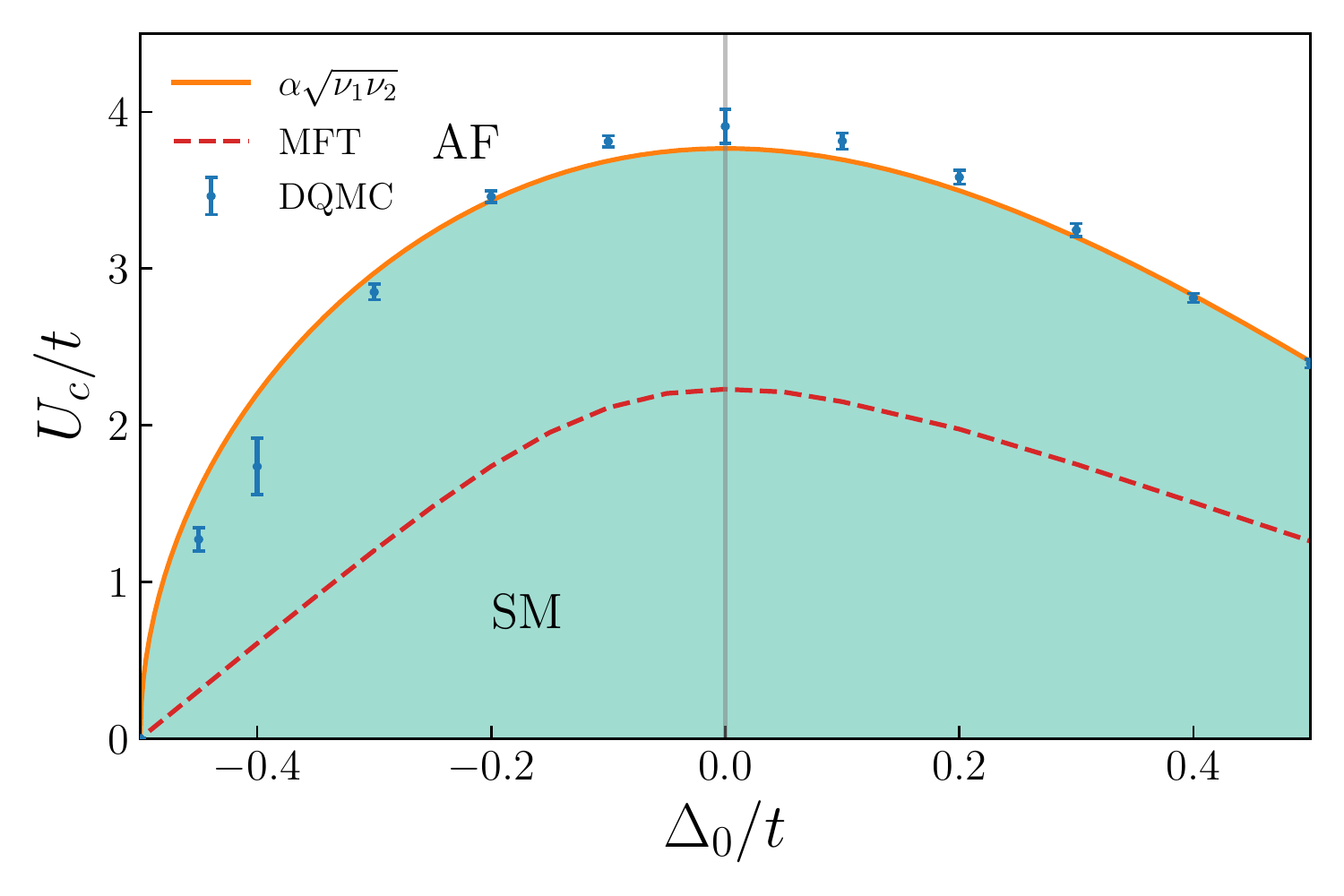} \caption{The phase diagram in the $(\Delta_0,U_c)$ plane. The dashed line represents the mean-field boundary, which underestimates the critical interaction. The DQMC critical values can be well fitted using the ansatz $U_c=\alpha\sqrt{\nu_1\nu_2}$, implying the critical interaction is proportional to the geometric average of the two velocities of the Dirac cones.}
\label{fig3}
\end{figure}

\paragraph{\color{blue}DQMC approach.}
In the DQMC approach,
Eq.~\eqref{eq1} at finite interactions is solved numerically, where one decouples the on-site interaction term through the introduction of an auxiliary Hubbard-Stratonovich field, which is then integrated out stochastically~\cite{PhysRevD.24.2278,PhysRevLett.51.1900,PhysRevB.31.4403,PhysRevB.40.506}. The only errors are those associated with the statistical sampling, the finite spatial lattice size, and the inverse temperature discretization. These errors can be well controlled in the sense that they can be systematically reduced as needed, and further eliminated by appropriate extrapolations. At half filling, the simulation is free of the sign problem, due to the presence of particle-hole symmetry~\cite{Loh1990,Troyer2005,Iglovikov2015,Li2015241117}. Thus we can access low enough temperatures, necessary to determine the ground-state properties on finite-size lattices. In our calculation, we use the inverse temperature $\beta=20$ and its discretization $\Delta\tau=0.1$. The lattice has totally $N=6\times L \times L$ sites, with $L$ up to $8$.

Applying DQMC to our problem, the antiferromagnetic order is characterized by the staggered structure factor~\cite{varney2009} with
\begin{equation}
  S^{z}_\text{AF}=\frac{1}{N}\sum_{i,j}\textrm{sgn}(i,j)\langle S^z_i S^z_{j}\rangle,
\end{equation}
where $\textrm{sgn}(i,j)=+(-)$ when $i,j$ belong to the same (opposite) sublattice.
Since the Hubbard model in Eq.~(\ref{eq1}) preserves the spin $SU(2)$ symmetry, so the spin-spin correlations of the three spin components are identical, and we only consider the $z$-component here. A related physical quantity of
interest here is the sublattice magnetization, which is given by $m_s=\sqrt{S^z_\text{AF}/N}$.

Figure~2(b) shows calculated $m_s^2$  as a function of $U$ on a $L=6$ lattice for various negative values of $\Delta$.
At $\Delta = 0$, it is known that AF order exists when $U$ exceeds $U_c = 3.86$~\cite{meng2010quantum,sorella2012absence,PhysRevX.3.031010,PhysRevX.6.011029,PhysRevB.72.085123,PhysRevB.91.165108}. In the $\Delta_0 = -0.5$ limit, the geometry corresponds to the $1/6$-depleted honeycomb lattice, where AF order is expected to exist for all $U > 0$, due to the existence of flat band. The behavior of $m_s^2$ versus $U$ is qualitatively similar for different values of $\Delta$: $m_s^2$ increases continuously with $U$, thus the semimetal-AF transition is of second-order nature. In addition, as $\Delta$ increases, the curves shift to the weak-interaction side, which results from the decrease of the critical interaction $U_c$ as the absolute value of $\Delta$ increases. These confirm the observation from the mean field approach.

To gain additional insight into the behavior of AF order, it is useful to examine the equal-time real space spin-spin correlation function $c({\bm r}) = \langle (n_{j+{\bm r}\uparrow}-n_{j+{\bm r}\downarrow} )(n_{j\uparrow}-n_{j\downarrow} )\rangle$. Figure~4 shows $c({\bm r})$ for $\Delta=-0.3$ at $U/t=2, 4$ on a  $L=6$ lattice. The origin is placed on the $1$st site of the unit cell at $(0, 0)$, and ${\bm r}$ runs along a triangular path (see Fig.4(a)). Since $U/t=2$ is below the critical interaction ($U_c/t=2.85$, see the following finite-size scaling), the values of the correlation at large distance are almost zero. On the other hand, the correlation length becomes comparable to the system size for the case of $U/t=4$. $c({\bm r})$ has a robust persistence at large distance, and its sign is consistent with AF order. This behavior is consistent with the fact that $U/t=4$ is above the critical point, and there exists AF order in the system. The sites in the lattice can be classified into three categories according to the number of the weakened bonds connected. $c({\bm r})$ varies among the different kinds of pairs of sites, which is most pronounced in the large-$U$ limit. As shown in Fig.~4(d), it increases with the total number of weakened bonds connected to the two sites in each pair, exhibiting an interesting behavior: The less the site is connected to the lattice, the stronger it is correlated to other sites.

\begin{figure}[htbp]
\centering \includegraphics[width=8.5cm]{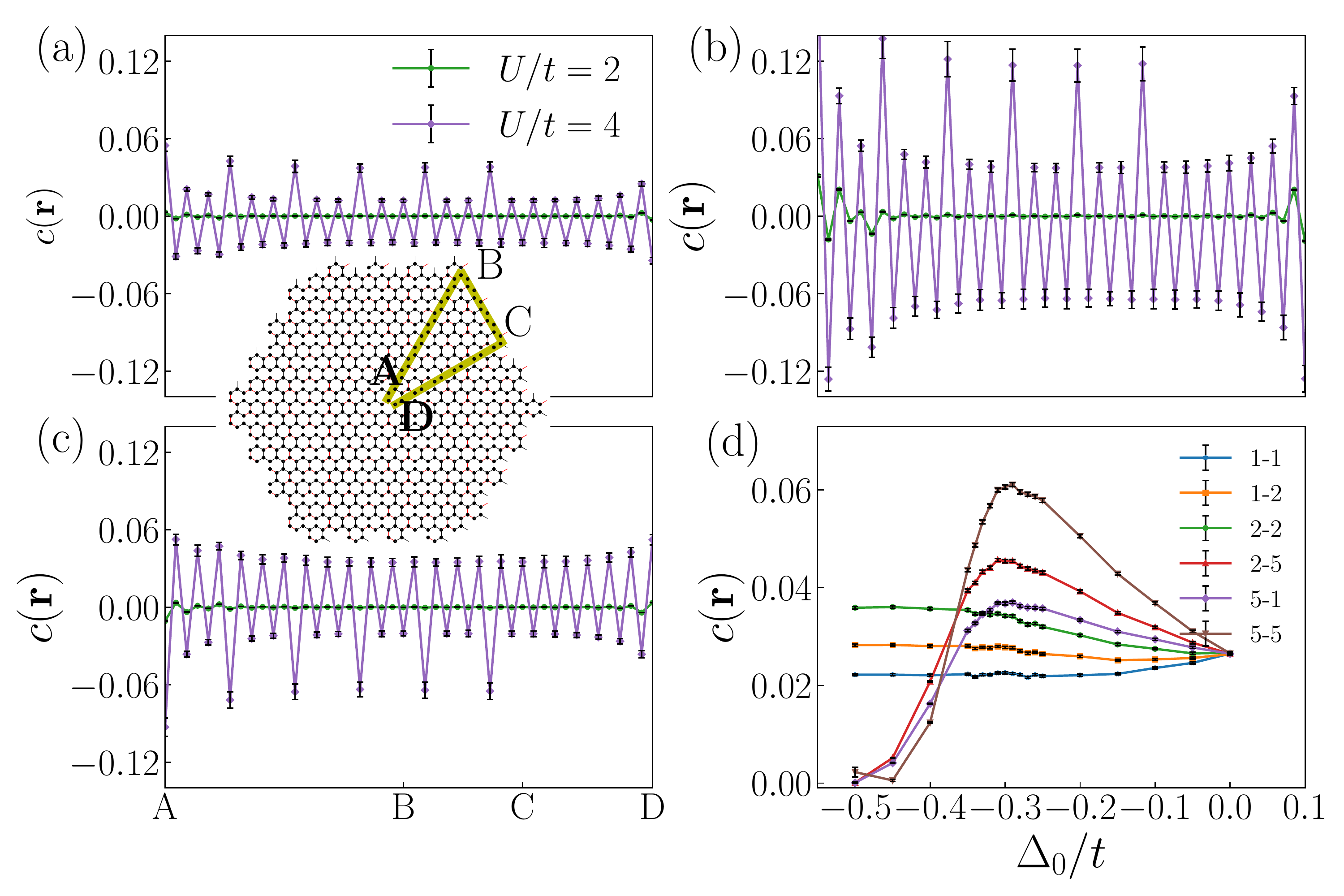} \caption{The spin correlations along the high-symmetry paths [see inset of (a) and (c)]. The origin is placed on the (a) 1st-, (b) 2nd-, and (c) 5th-sites of the unit cell at $(0, 0)$, which are representatives of the different ones in the unit cell. Here the anisotropic parameter is $\Delta_0/t=-0.3$. (d) The spin-spin correlations as a function of $\Delta_0$ for the largest distance on a $L=20$ lattice in the large-$U$ limit.}
\label{fig4}
\end{figure}

The behavior of $S^z_\text{AF}$ indicates that AF order may develop at a decreased critical interaction in
the presence of $Y$-shaped bond modulation in going from the pristine honeycomb lattice to the limiting cases of $\Delta=-0.5, 1$. We then use finite-size scaling to analyze quantitatively the position of the critical point in the thermodynamic limit. The square of the order parameter is given by $S^z_\text{AF} /N$ in the $1/L \rightarrow 0$ limit. These extrapolated values are shown in the phase diagram Fig.~3. As a function of $\Delta$, the critical interaction strength continuously decreases from $U_c/t=3.869$ to zero. Besides, we use the ansatz $U_c = \alpha\sqrt{v_1 v_2}$ to fit the boundary, and find that the critical values are well fitted with $\alpha=2.51$. This shows that
the critical interaction is proportional to the geometric average of the two velocities of birefringent Dirac fermions. This relation is not known in previous studies on the uniform honeycomb-lattice and $\pi$-flux Hubbard models, because the velocity there keeps a constant~\cite{PhysRevB.91.165108}. Here, the birefringent setup allows a continuous tuning of the velocities, thus providing understanding on the relationship between the critical interaction and the Dirac velocities.
It is also noted that when the value of $\Delta$ is large enough, the critical interaction can be well below the actual $U$ value in graphene, resulting in the emergence of AF order.

We also perform a finite-size scaling analysis based on the usual scaling formula~\cite{assaad2013},
\begin{equation}
  S_\text{AF}^{z}=L^{2-2\beta/\nu} F[L^{1/\nu}(U-U_c)],
\end{equation}
where $\beta$ is the order parameter exponent, and $\nu$ is the correlation length exponent.
The semimetal-AF transition is expected to belong to the Gross-Neveu-Heisenberg universality class~\cite{PhysRevLett.97.146401,PhysRevB.79.085116}. The previous DQMC studies reported $\nu=1.02$ and $\beta=0.76$~\cite{PhysRevX.6.011029}.
Together with the critical interaction determined by the finite-size scaling above, we scale $S_\text{AF}^{z}$ at different lattice sizes according to the above formula. As shown in Fig.~\ref{fig5}(b), the data collapse is pretty good, thus confirming the universality class of the phase transition here is unchanged by the $Y$-shaped Kekul\'e distortion.

In addition, we have further checked the single-particle gap by extracting the spectral function and the density of states from the imaginary-time Green function using analytic continuation~\cite{mynote}. It shows that the AF transition is always accompanied by a charge-gap opening, indicating that the system becomes a AF Mott insulator above $U_c$.

\begin{figure}[t]
\centering \includegraphics[width=8.5cm]{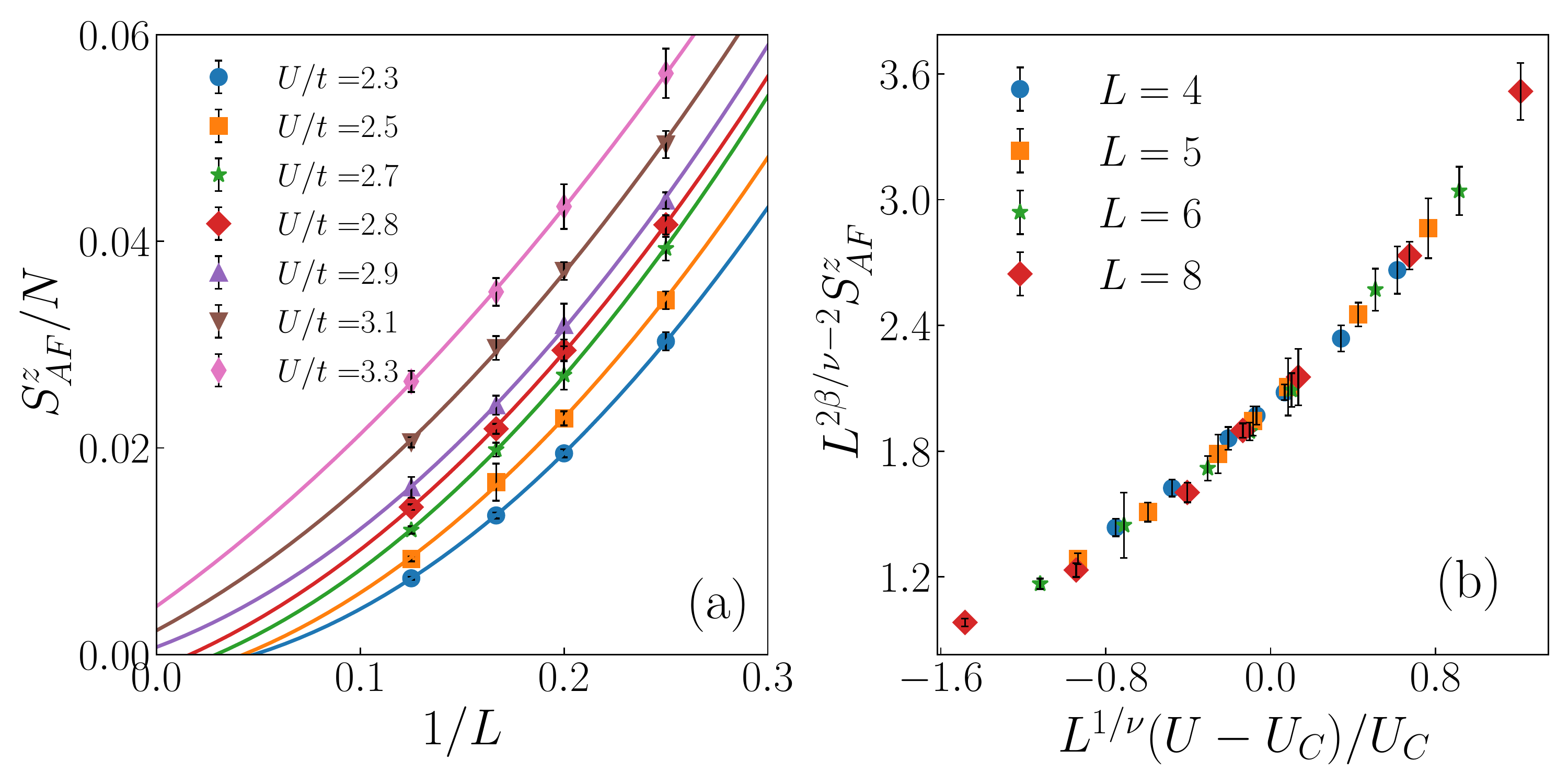} \caption{(a) Extrapolation of the structure factor $S^z_{AF}/N$ near the transition point. The solid lines are least-squares fits to the polynomial form of $1/L$. The value in the thermodynamic limit becomes finite at some interaction between $U/t=2.8$ and $2.9$, thus the critical interaction is estimated to be $U_c/t=2.85\pm 0.05$. (b) The data collapse using the critical exponets of the Gross-Neveu-Heisenberg universality class and the critical interaction determined in (a). Here the anisotropic parameter is $\Delta_0/t=-0.3$.}
\label{fig5}
\end{figure}

\paragraph{\color{blue} Discussion.}
We have applied the mean-field theory and DQMC simulations to study the Hubbard
model on a honeycomb lattice with a $Y$-shaped distortion. Both approaches reveal that AF order develops above a critical interaction, and the critical interaction decreases monotonically with the distortion parameter, and scales with the geometric average of the two velocities of the birefringent Dirac fermions. We find that the quantum criticality of the continuous AF transition is unchanged by the distortion and still belongs to the Gross-Neveu-Heisenberg universality class.

The fact that the $Y$-shaped Kekul\'e distortion can continuously tune the quantum critical point is of great significance in graphene research and applications. When the critical point is put below the actual value of $U$ in graphene~\cite{mynote1,tang2018role}, the long-sought bulk magnetism in graphene can be realized. Recent experiments~\cite{gutierrez2016imaging,helin2022} have demonstrated the realization of $Y$-shaped distortion in graphene.
For graphene on Cu(111), the distortion is likely due to regular copper vacancies which lead to a vertical shift of the central carbon atom of each $Y$-shape texture~\cite{gutierrez2016imaging}. Our result indicates that when the shift is large enough, AF order should be spontaneously generated in the graphene layer.
Thus, our results not only deepen our understanding of the semimetal-AF transitions in Dirac fermion systems, but also provide a feasible approach to induce magnetism in graphene.


\paragraph{\color{blue}Acknowledgments.}
C.W., W.H. and H.G. acknowledge support from the National Natural Science Foundation of China (NSFC) grant Nos.~11774019 and 12074022, the NSAF grant in NSFC with grant No. U1930402, the Fundamental Research
Funds for the Central Universities and the HPC resources
at Beihang University. X.F. and S.A.Y are supported by the Singapore MOE AcRF Tier 2 (MOE-T2EP50220-0011).
S.F. is supported by the National Key Research and
Development Program of China under Grant No. 2021YFA1401803,
and NSFC under Grant Nos. 11974051 and 11734002.

\bibliographystyle{apsrev4-2}
\bibliography{yshape_v0909}

\begin{thebibliography}{53}
\bibitem [{myn({\natexlab{a}})}]{mynote1}%
  \BibitemOpen
  \href@noop {} {} ({\natexlab{a}}),\ \bibinfo {note} {estimates for the
  graphene Hubbard $U$ vary significantly in the literature. Here we use
  $U/t=3.0$, which is obtained using the quantum chemistry Pariser-Parr-Pople
  method in Ref.[28].}\BibitemShut {Stop}%
\end{thebibliography}


\clearpage

\renewcommand{\theequation}{S\arabic{equation}}
\setcounter{equation}{0}


\begin{center}

{\large \bf Supplementary Materials:
 \\Emergent antiferromagnetism in a $Y$-shaped Kekul\'e graphene}\\

\vspace{0.3cm}

\end{center}

\vspace{0.6cm}

\beginsupplement

\title{Supplementary Materials:
 \\The bilayer Hubbard model: analysis based on the fermionic sign problem}
\maketitle

\renewcommand{\thefigure}{A\arabic{figure}}

\setcounter{figure}{0}

In these Supplementary materials we present the details of the band structures, the properties of the single-particle excitations, and more results for the spin correlations.

\section{The band structures at various values of $\Delta_0$}

In momentum space, the $U=0$ Hamiltonian is given by
\begin{align}
\mathcal{H}_{0}({\mathbf{k}})=\left(\begin{array}{cc}
0 & h_{12} ({\mathbf{k}}) \\
h_{12}^{\dagger} ({\mathbf{k}}) & 0
\end{array}\right),
\end{align}
where
\begin{align}
h_{12} ({\mathbf{k}})=-\left(\begin{array}{ccc}
\Delta_{-} & \Delta_{-}e^{i{\bf k}\cdot {\bf a}_1} & \Delta_{-} \\
\Delta_{-} & \Delta_{-}  & \Delta_{-}e^{i{\bf k}\cdot ({\bf a}_2-{\bf a}_1)}\\
\Delta_{+}e^{i{\bf k}\cdot {\bf a}_2} & \Delta_{+}  & \Delta_{+}
\end{array}\right),
\end{align}
with $\Delta_{-}=1-\Delta_0$, $\Delta_{+}=1+2\Delta_0$, and the lattice constants ${\mathbf{a}}_1=3(\sqrt{3}/2,-1/2) , {\mathbf{a}}_2=3(\sqrt{3}/2,1/2)$. The whole spectrum has six dispersive bands, which can be obtained by directly diagonalizing the Hamiltonian in Eq.(1). The maximum value of the eigenenergy is at $\Gamma$ point, which is $6t\sqrt{2\Delta_0^2+1}$. Although the bandwidth varies with $\Delta_0$, the way to parameterize the strong and weak bonds used in the main text minimizes the variation of the bandwidth as $\Delta_0$ changes.
Another advantage of this kind of choice is that the bandwidth is independent of the sign of $\Delta_0$.

\begin{figure}[htbp]
\centering \includegraphics[width=9cm]{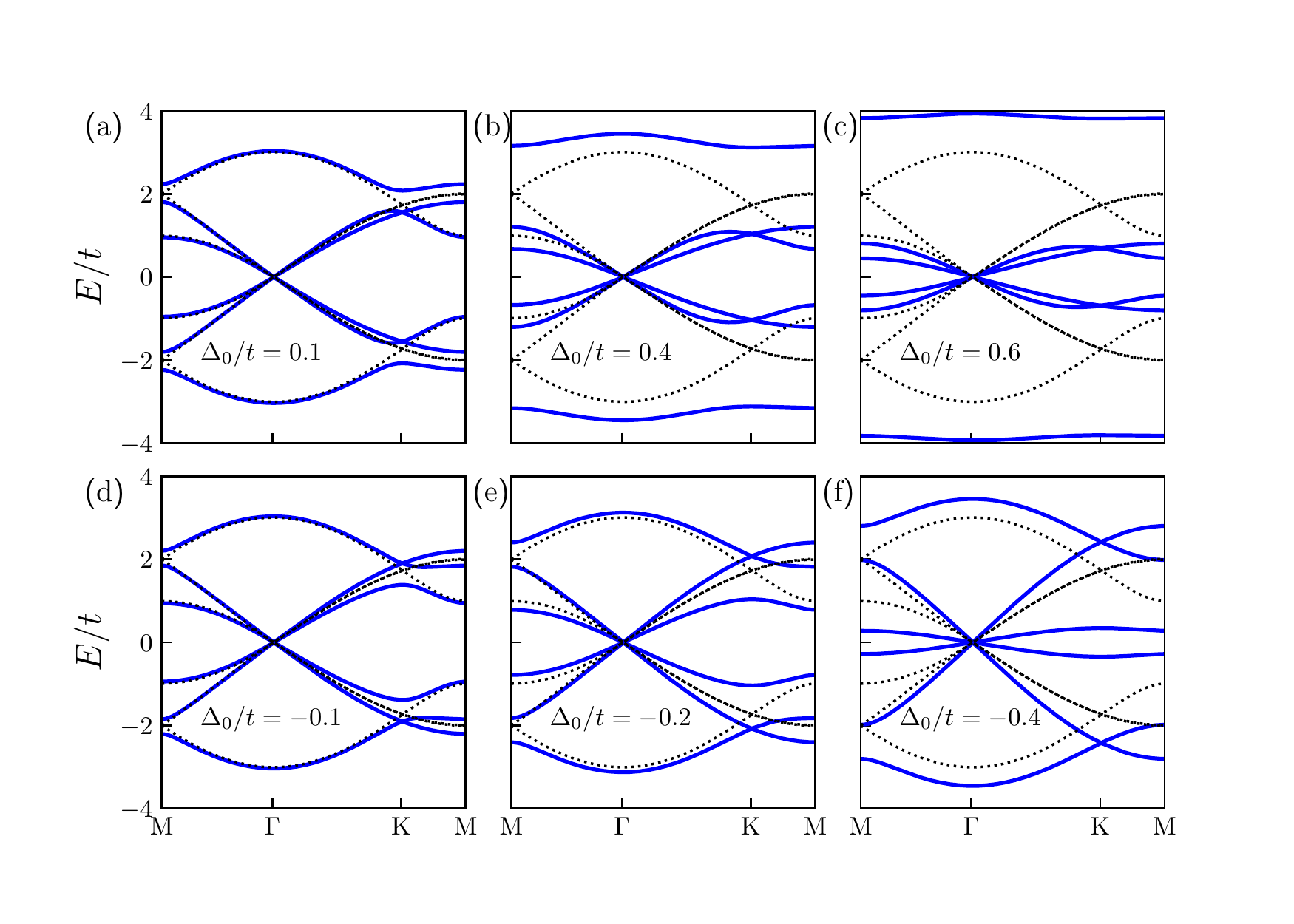} \caption{The band structures at: (a) $\Delta_0/t=0.1$, (b) $\Delta_0/t=0.4$, (c) $\Delta_0/t=0.8$,
 (d) $\Delta_0/t=-0.1$, (e) $\Delta_0/t=-0.2$, (f) $\Delta_0/t=-0.4$. The dotted lines in each figure represent the band structure of graphene, i.e., the $\Delta_0/t=0$ case, which are plotted for the purpose of comparison.}
\label{afig1}
\end{figure}

When $\Delta_0$ varies from $0$ to $-0.5$, the bandwidth of the two bands near the Fermi energy decreases continuously, and becomes vanished at $\Delta_0/t=-0.5$. The evolution of the bandwidth is similar in changing $\Delta_0$ from $0$ to $1$. The difference is that the band-flatting process involves the four bands near the Fermi energy here.

\begin{figure}[htbp]
\centering \includegraphics[width=6.5cm]{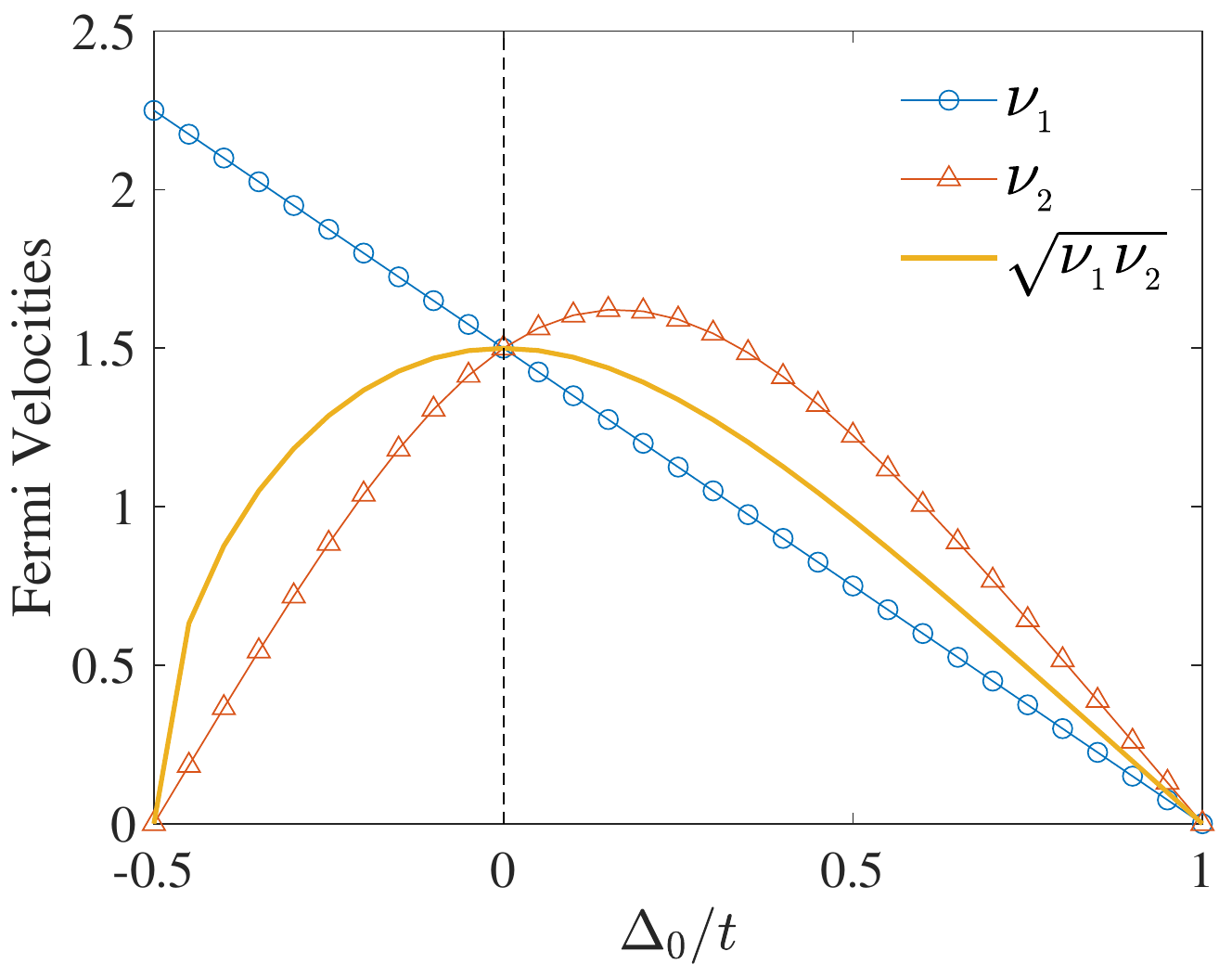} \caption{The Fermi velocities $\nu_1,\nu_2$ and their geometric average $\sqrt{\nu_1\nu_2}$ as a function of $\Delta_0$.}
\label{afig2}
\end{figure}

\subsection{The flat-band states at $\Delta_0/t=-0.5$}
We have $\Delta_{+}=0$ at $\Delta_0/t=-0.5$, when the Hamiltonian in Eq.(1) can be solved analytically. Generally there are two zero-energy states at each momentum. As has been stated in the main text, one of them is totally localized on the $5$th site, which is completely isolated from the lattice. The wave function of the other one is $\psi_0({\bf k})=\lambda[0,0,0,f_1({\bf k}),f_2({\bf k}),1]^{T}$, where $f_1({\bf k})=(1-e^{i{\bf k}\cdot {\bf a}_2})/(e^{i{\bf k}\cdot {\bf a}_1}-1)$, $f_2({\bf k})=[1-e^{i{\bf k}\cdot ({\bf a}_2-{\bf a}_1})]/(e^{i{\bf k}\cdot {\bf a}_1}-1)$, and $\lambda=1/\sqrt{1+|f_1({\bf k})|^2+|f_2({\bf k})|^2}$. Hence this flat-band state distributes only on the $2$nd, $4$th, and $6$th sites of the unit cell, each of which is connected by one red bond.

In real space, the above zero-energy states associated with the connected lattice can be constructed within each triple hexagon centered at the isolated site. Each of such states only distribute on the six sites connected by two bonds.  The weights among the sites are equal, and the phases of the wave function alternate between $1,-1$ along the edge of the triple hexagon.

\begin{figure}[htbp]
\centering \includegraphics[width=6.5cm]{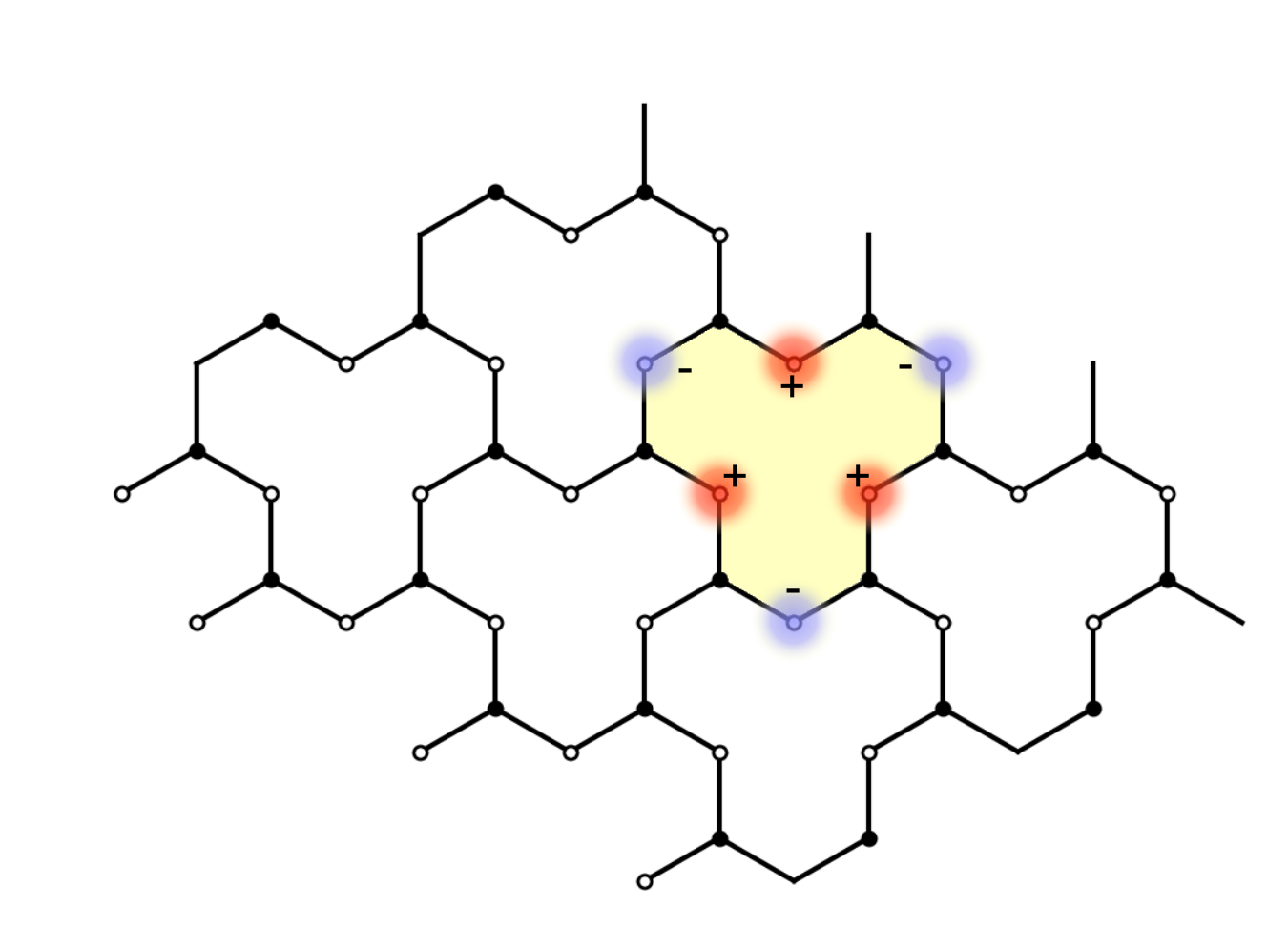} \caption{The wave function of one of the flat-band states at $\Delta_0/t=-0.5$. Each of such
states is localized within the triple hexagon, and only distribute on the six sites connected by two bonds. The weights among the above sites are equal, and the phases of the wave function alternate between $1,-1$ along the edge of the triple hexagon.}
\label{afig3}
\end{figure}

\subsection{The flat-band states at $\Delta_0/t=1$}

At $\Delta_0/t=1$, the lattice is decoupled into isolated sites and four-pointed stars. Four of the six eigenvalues at each ${\bf k}$ are zero-energy states. Two of them are completely localized on the isolated sites, and the other two are within the four-pointed stars, for which the Hamiltonian writes as
\begin{align}
H=-t_{+}\left(\begin{array}{cccc}
0 & 1 & 1 & 1 \\
1 & 0 & 0 & 0 \\
1 & 0 & 0 & 0 \\
1 & 0 & 0 & 0
\end{array}\right).
\end{align}
Its eigenvalues are $\pm \sqrt{3}t_{+},0,0$. Two linearly independent eigenstates for the zero-energy states can be constructed as: $\psi_1=\frac{1}{\sqrt{2}}[0,1,-1,0]^{T}$, $\psi_2=\frac{1}{\sqrt{2}}[0,1,0,-1]^{T}$. The distributions of the above wave functions are only on the surrounding three sites, each of which is only connected by one bond.

\subsection{Deducing of the low-energy effective Hamiltonian}

At ${\bf k}=(0,0)$, the eigenvectors corresponding to the four zero-energy states of the Dirac points forms the projection matrix,
\begin{equation}
	P = \begin{pmatrix}
		0&0&-\frac{1}{\sqrt{2}}&\frac{-1-2\Delta_0}{\sqrt{-6+12\Delta_0^2}}\\
		0&0& \frac{1}{\sqrt{2}}&\frac{-1-2\Delta_0}{\sqrt{-6+12\Delta_0^2}}\\
		0&0&                0  &\frac{\sqrt{2}(1-\Delta_0)}{\sqrt{3+6\Delta_0}}\\
		-\frac{1}{\sqrt{2}}&-\frac{1}{\sqrt{6}}&0&0\\
		0                  &\sqrt{\frac{2}{3}}&0&0\\
		\frac{1}{\sqrt{2}}&-\frac{1}{\sqrt{6}}&0&0\\
	\end{pmatrix}.
\end{equation}
Then the effective low-energy Hamiltonian is obtained by
\begin{equation}
	H_{eff}({\bf k}) = P^T {\cal H}_0({\bf k}) P
\end{equation}
which is
\begin{align}
\mathcal{H}_{eff}({\mathbf{k}})&=\left(\begin{array}{cc}
0 & f ({\mathbf{k}}) \\
f^{\dagger} ({\mathbf{k}}) & 0
\end{array}\right), \\
f({\mathbf{k}})&=-\frac{3\mathbb{I}}{2}(1-\Delta_0)t\left(\begin{array}{cc}
k_y & \alpha k_x \\
-k_x &\alpha k_y
\end{array}\right),
\end{align}
where the anisotropic factor is $\alpha=\frac{1+2\Delta_0}{\sqrt{1+2\Delta^2_0}}$.

\section{The single-particle excitation}

To investigate the single-particle gap, we use analytic continuation to extract the spectral function from the  imaginary-time dependent Green function $G(\tau,{\bf k})=\langle c_{\bf k}(\tau)c_{\bf k}^{\dagger}(0)\rangle$,
\begin{align}
G(\tau,{\bf k})=\frac{1}{\pi}\int_{-\infty}^{\infty} d\omega \frac{e^{-\tau\omega}}{1+e^{-\beta\omega}}A({\bf k},\omega).
\end{align}
Then the density of states, $N(\omega)$, can be directly calculated,
\begin{align}
N(\omega)=\int d k A(k, \omega).
\end{align}

$N(\omega)$ shown in Fig. 4 characterizes a metal-insulator transition driven by the Hubbard interaction $U$. The density of states is finite at $\omega=0$ for small $U$. When $U$ is large enough, it is zero in a finite region near $\omega=0$, which corresponds to the size of a charge gap. Compared to the $\Delta_0=0$ case, the gap opening occurs at smaller $U$ for $\Delta_0=-0.4$. The critical interactions estimated qualitatively are consistent with those determined from the antiferromagnetic (AF) transition. This implies the AF transition is accompanied by a charge-gap opening, suggesting the system is a AF Mott insulator above the critical point.

The spectral function $A({\bf k},\omega)$ counts single-particle excitations at a given momentum and energy, thus allows us to locate the point in the Brillouin zone where the single-particle gap is minimum. As shown in Fig.5, the gap is minimum at the $\Gamma$ point, thus its opening occurs at the Dirac points.

\begin{figure}[htbp]
\centering \includegraphics[width=8.5cm]{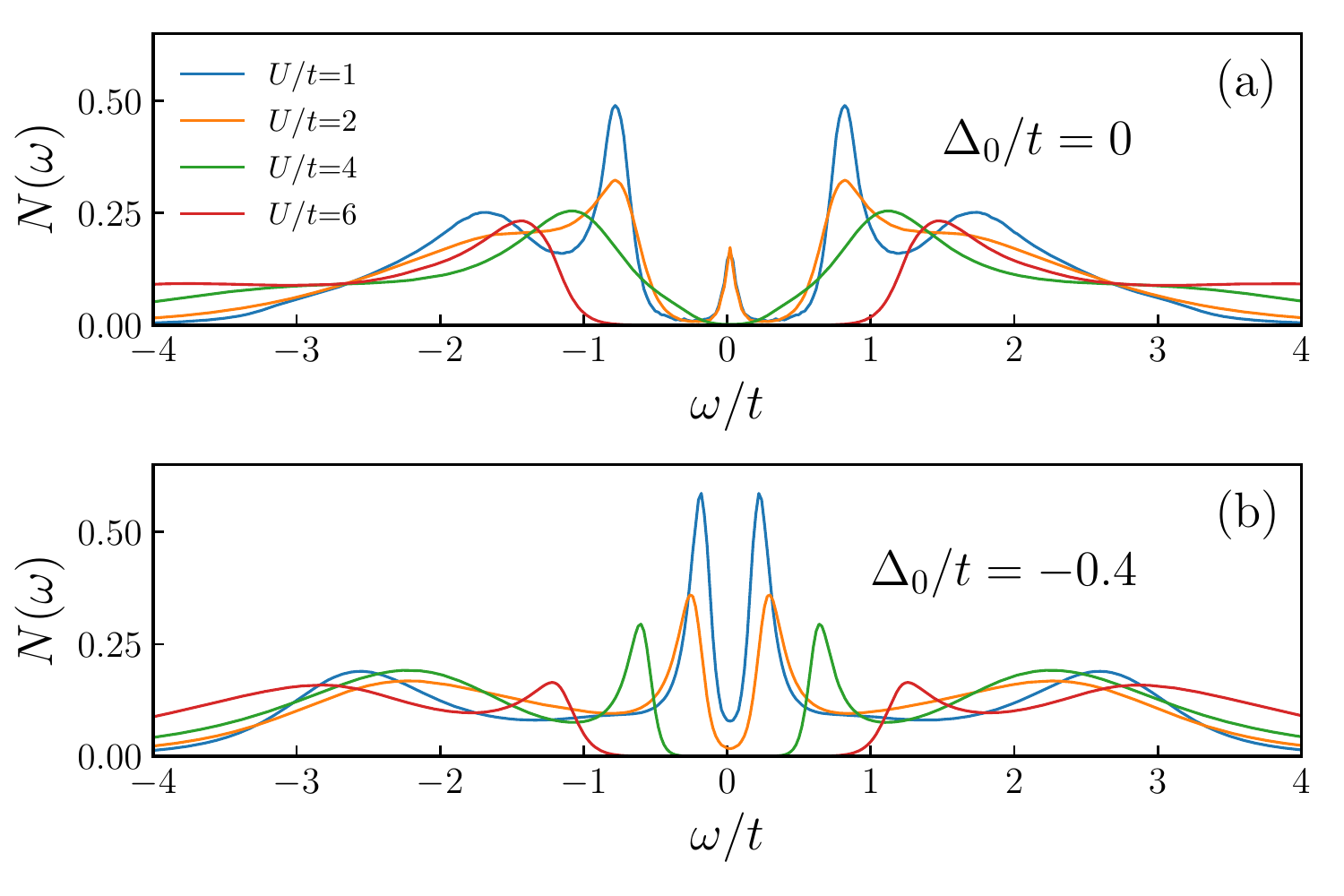} \caption{The density of states $N(\omega)$ at various values of $U$: (a) $\Delta_0/t=0$ (b) $\Delta_0/t=-0.4$.}
\label{afig4}
\end{figure}

\begin{figure}[htbp]
\centering \includegraphics[width=8.5cm]{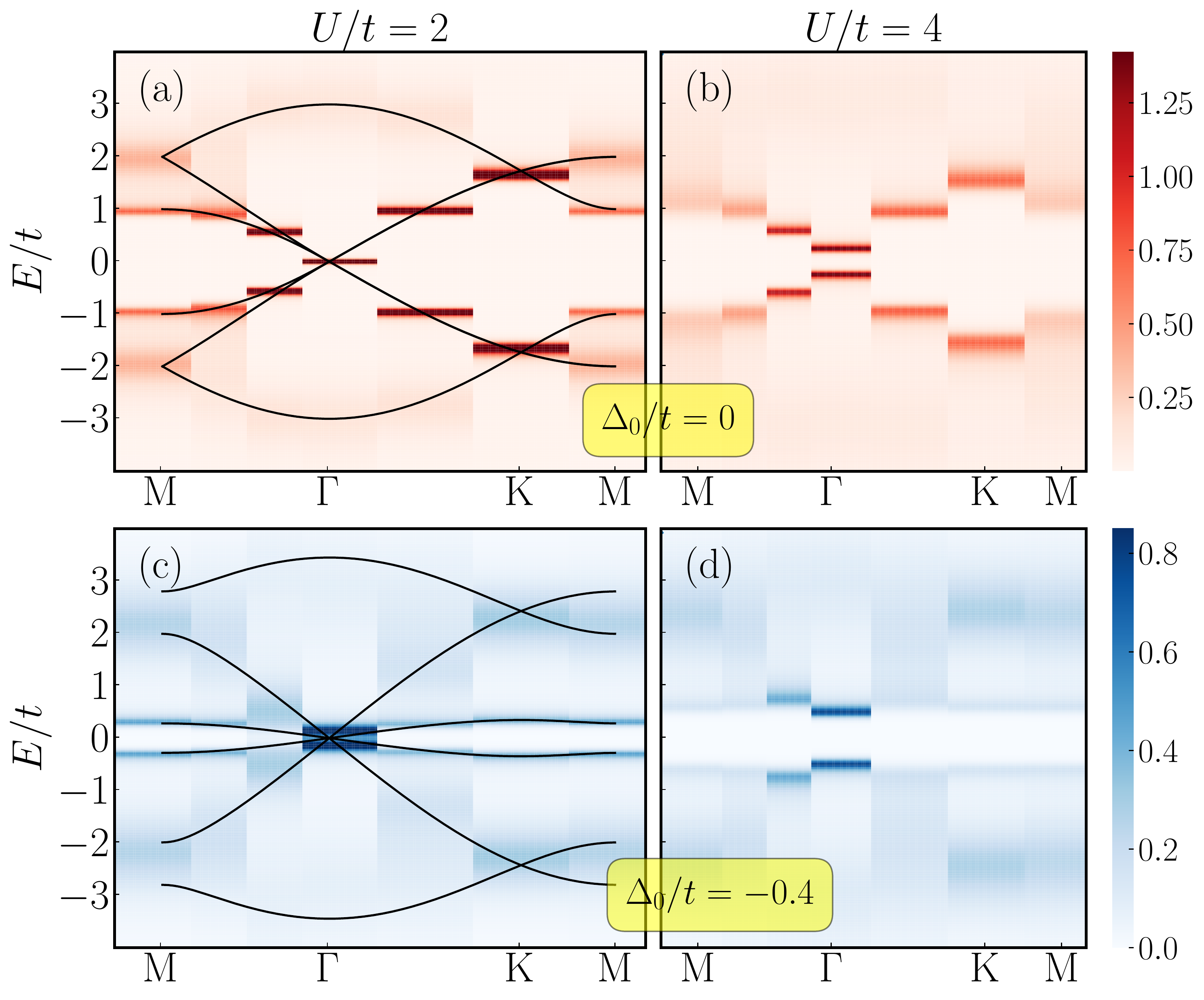} \caption{The single-particle spectral function: (a) $\Delta_0/t=0$, $U/t=2$; (b) $\Delta_0/t=0$, $U/t=4$; (c) $\Delta_0/t=-0.4$, $U/t=2$; (d) $\Delta_0/t=-0.4$, $U/t=4$. Here the lattice size is $L=6$, and the inverse temperature is $\beta=20$.}
\label{afig5}
\end{figure}

\section{The spin correlations}

Figure 6(b) shows the AF structure factor as a function of $U$ for various values of positive $\Delta_0$. For small $\Delta_0$, $S_{AF}^z/N$ increases continuously with $U$, implying the existence of a continuous SM-AF transition. In addition, as $\Delta_0$ increases, the curves shift to the weak-interaction side, thus the critical interaction $U_c$ should decrease as $\Delta_0$ increases. The above behavior is consistent with the phase diagram of the main text. The case of $\Delta_0/t=0.5$ is special, where $S_{AF}^z/N$ has a clear drop at sufficient large $U$. It should result from the collapse of the AF order at strong $U$ for large $\Delta_0$, where an AF to $Y$-dimer transition occurs.

\begin{figure}[htbp]
\centering \includegraphics[width=8.5cm]{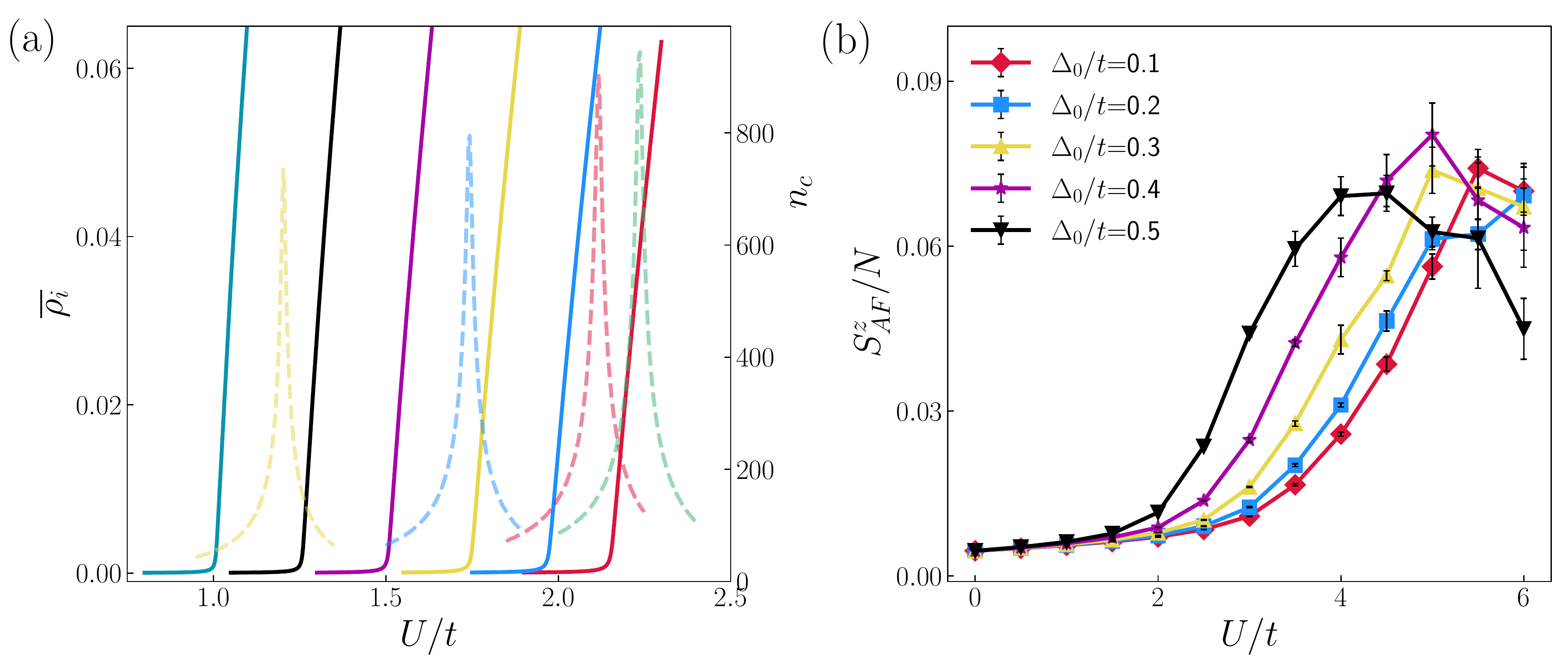} \caption{(a) The mean-field order parameter and the cycling time $n_c$ to get convergent solution in the self-consistent process as a function of $U$ for various values of positive $\Delta_0$. The sharp peak in the $n_c$ curve can steadily determine the transition point. (b) The AF structure factor obtained by DQMC as a function of $U$ on a lattice with the size $L=6$.}
\label{afig6}
\end{figure}

\begin{figure}[htbp]
\centering \includegraphics[width=8.5cm]{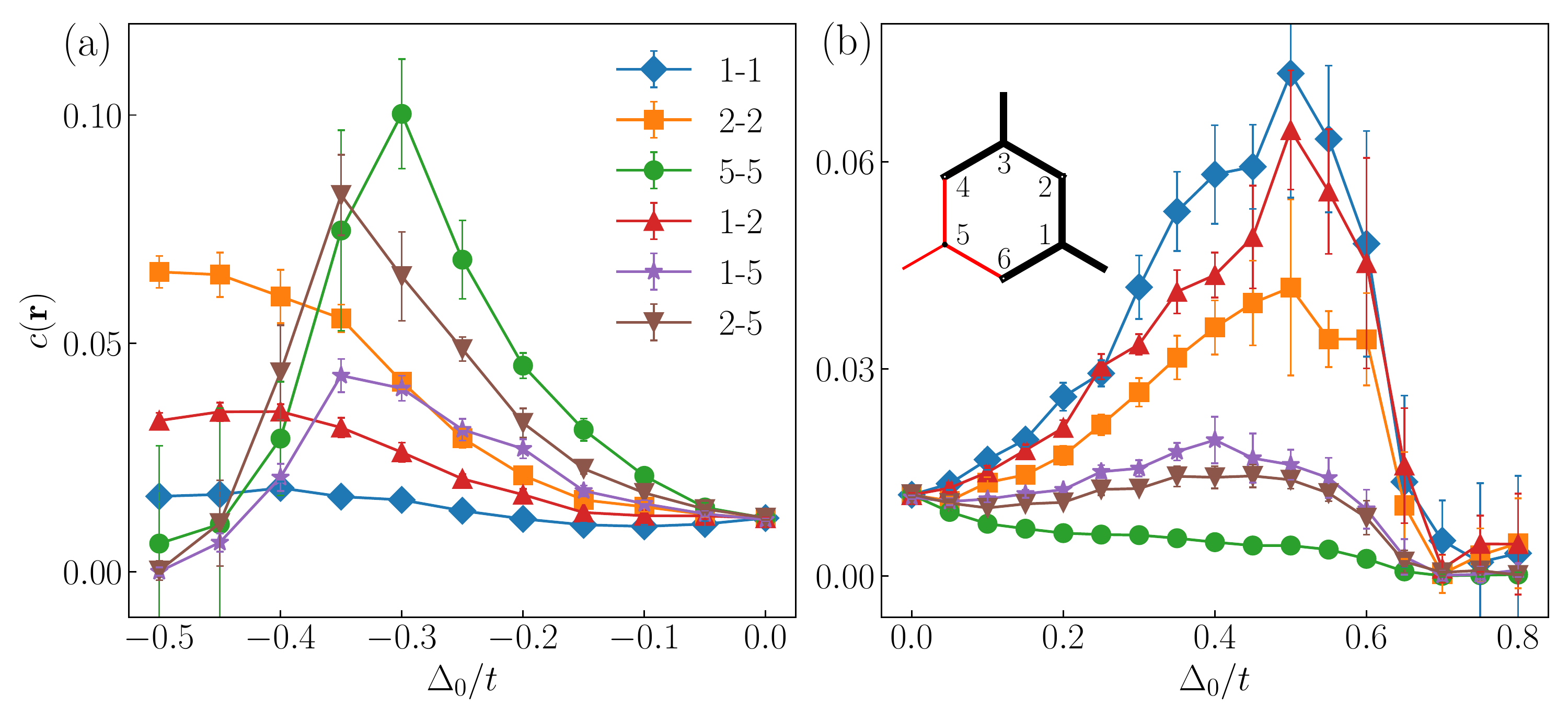} \caption{The spin-spin correlations as a function of $\Delta_0$ between
the sites with the same and different indexes of the unit cells in the regions: (a) $\Delta_0<0$; (b) $\Delta_0>0$. Here ${\bf r}$ takes the largest distance in a $L=6$ lattice, and the strength of the interaction is $U/t=6$.}
\label{afig7}
\end{figure}

Figure \ref{afig7} plots the spin-spin correlation function between different kinds of pairs of sites as a function of $\Delta_0$. The largest distance between the two sites in each kind of pair is considered, which can represent the long-range spin correlation. Figure \ref{afig7}(a) shows $c({\bf r})$ in the region $\Delta_0<0$. The $1$st, $2$nd, and $5$th sites in each unit cell are connected by zero, one, and three weakened bonds, respectively. The values of the spin correlations have the following relation: $c_{5-5}({\bf r})>c_{2-2}({\bf r})>c_{1-1}({\bf r})$. For $c({\bf r})$ between the sites with different indexes, the relation of the corresponding values is: $c_{2-5}({\bf r})>c_{1-5}({\bf r})>c_{1-2}({\bf r})$. Hence when the absolute value of $\Delta_0$ is small, the value of $c({\bf r})$ increases with the total number of weakened
bonds connected to the two sites in each pair, which remains valid when all the values in Fig.\ref{afig7}(a) are compared. This suggests the spin correlation between two sites can be strengthened by locally weakening the bonds connecting them. As $\Delta_0$ increases, the values of the spin correlations without the $5$th index involved continuously increases and tends to constants in the $\Delta_0/t=-0.5$ limit. The sequence of the values keeps as $c_{2-2}({\bf r})>c_{2-1}({\bf r})>c_{1-1}({\bf r})$. In contrast, the spin correlations involving the $5$th site begin to decrease quickly from $\Delta_0/t\sim -0.3$, and becomes zero at $\Delta_0/t=-0.5$, which is expected since the $5$th site is completely depleted from the lattice in this limit. As demonstrated in Fig.\ref{afig7}(b), the case with $\Delta_0>0$ is similar, except that all values begins to drop at $\Delta_0/t \sim 0.5$, and becomes zero at $\Delta_0/t\sim 0.7$. Here the vanishment of the spin correlations at extremely large $\Delta_0$ is due to the occurrence of an AF to $Y$-dimer transition.

\begin{figure}[htbp]
\centering \includegraphics[width=6.5cm]{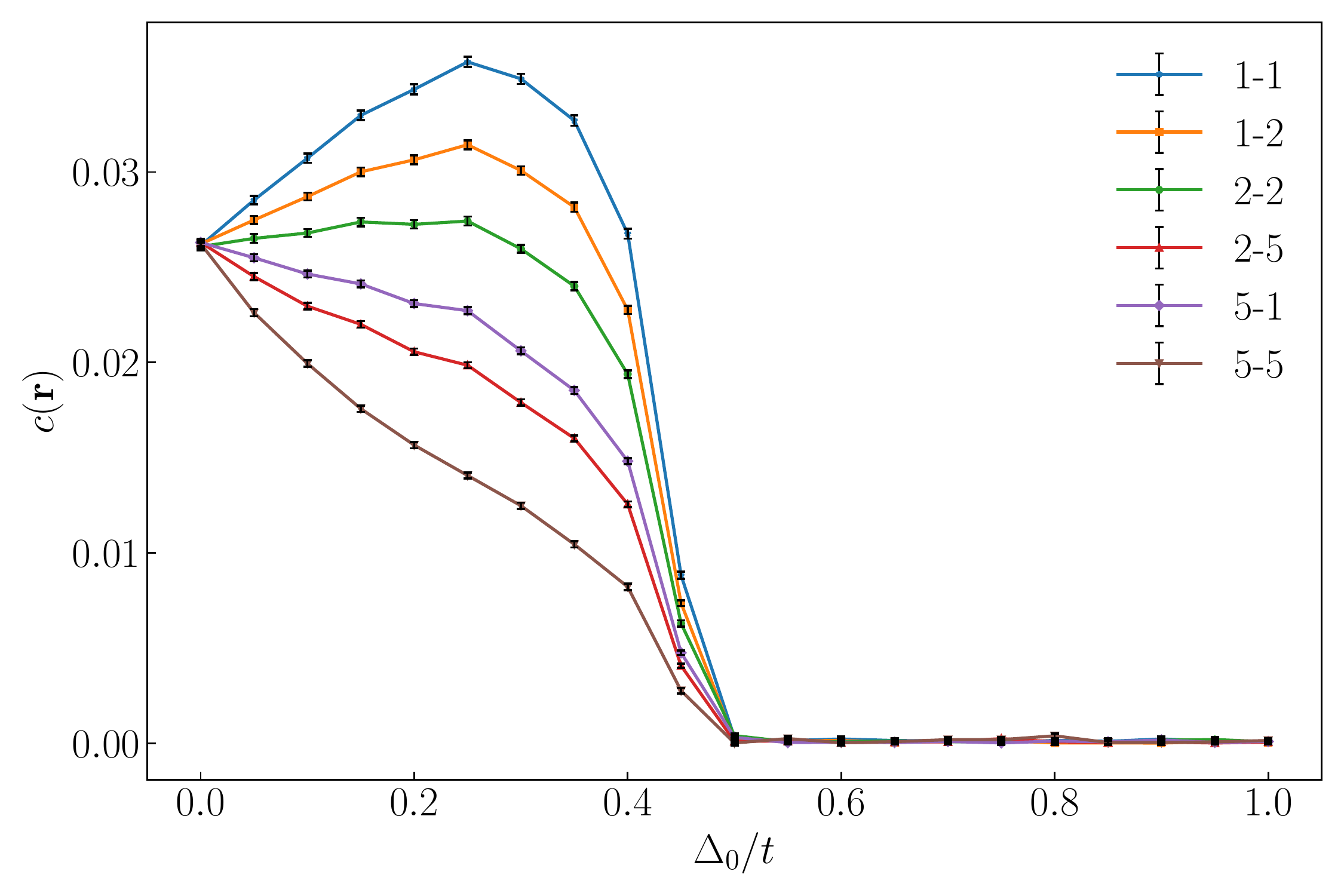} \caption{The spin-spin correlations as a function of positive $\Delta_0$ for the largest distance on a $L=20$ lattice in the large-$U$ limit.}
\label{afig8}
\end{figure}

In the large-$U$ limit, the double occupancy is completely eliminated, and the Hubbard model in Eq.(1) maps onto the following Heisenberg model\cite{charles1976},
\begin{align}
  {\cal H}=\sum_{\langle ij\rangle}J_{ij} {\bf S}_i\cdot {\bf S}_j,
\end{align}
where the exchange coupling is $J_{ij}=\frac{4t_{ij}^2}{U}$. Corresponding to the $Y$-shaped modulations of the hopping amplitudes, the value of $J_{ij}$ takes $J_1=4(1-\Delta_0)^2/U$ or $J_2=4(1+2\Delta_0)^2/U$. We set $J_1=1$ for $\Delta_0<0$ ($J_2=1$ for $\Delta_0>0$) as the energy scale, thus $J_2=(1+2\Delta_0)^2/(1-\Delta_0)^2$ ($J_1=(1-\Delta_0)^2/(1+2\Delta_0)^2$), which decreases monotonically as $\Delta_0$ changes from $0$ to $-0.5 (1)$.

Similar to the plot in Fig.4(d) of the main text, we here plot the spin-spin correlations at the largest distance on a $L=20$ lattice as a function of positive $\Delta_0$ in the large-$U$ limit. Now the $1$st, $2$nd, and $5$th sites in each unit cell are connected by there, one, and zero weakened bonds, respectively. The values of the spin correlations have the following relation: $c_{1-1}({\bf r})>c_{1-2}({\bf r})>c_{2-2}({\bf r})>c_{5-1}({\bf r})>c_{2-5}({\bf r})>c_{5-5}({\bf r})$. Hence the statement that the value of $c({\bf r})$ increases with the total number of weakened bonds connected to the two sites in each pair, remains valid. Different from the case of $\Delta_0<0$, the value of $c({\bf r})$ begins to be zero at $\Delta_0/t=0.5$, where an AF to $Y$-dimer transition occurs. Near the critical point, $c({\bf r})$ decreases rapidly, implying the AF order collapses quickly with increasing $\Delta_0$ here.


\end{document}